\documentclass[prl,twocolumn,superscriptaddress,showpacs,preprintnumbers]{revtex4}

\usepackage{amssymb}
\usepackage{amsmath}
\usepackage{graphicx}
\usepackage{dcolumn}
\usepackage{verbatim}
\usepackage{bm}

\newcommand{\be}{\begin{equation}}
\newcommand{\ee}{\end{equation}}

\begin{document}

\title{The differential information-geometry of quantum phase transitions}
\author{Paolo Zanardi}
\affiliation{Institute for Scientific Interchange, Villa Gualino, Viale
Settimio Severo 65, I-10133 Torino, Italy }
\affiliation{Department of Physics and Astronomy, University of Southern California Los
Angeles, CA 90089-0484 (USA) }
\author{Paolo Giorda}
\affiliation{Institute for Scientific Interchange, Villa Gualino, Viale
Settimio Severo 65, I-10133 Torino, Italy }
\author{Marco Cozzini}
\affiliation{Dipartimento di Fisica, Politecnico di Torino, Corso Duca degli
Abruzzi 24, I-10129 Torino, Italy}
\affiliation{Institute for Scientific Interchange, Villa Gualino, Viale
Settimio Severo 65, I-10133 Torino, Italy }
\date{\today}

\begin{abstract}
The manifold of coupling constants parametrizing a quantum Hamiltonian is
equipped with a natural Riemannian metric with an operational
distinguishability content.
We argue that the singularities of this metric are in correspondence with the
quantum phase transitions featured by the corresponding  system. This approach
provides a universal conceptual framework to study quantum critical phenomena 
which is differential-geometric and information-theoretic at the same  time. 
\end{abstract}

\pacs{03.65.Ud,05.70.Jk,05.45.Mt}
\maketitle
{\em Introduction}.--
Suppose you are given with a set of quantum states  associated to a family
of Hamiltonians  smoothly  depending on a set of parameters, e.g., coupling
constants. This parameter manifold -- that can include temperature in the case
the considered states are thermal ones -- is partitioned in regions
characterized by the fact that inside them  one can ``adiabatically" move from
one point to the other and no singularities in the expectation values of any
observables  are encountered. The boundaries between these  regular regions are
in turn associated to the non-analytic behaviour of some observable and are
referred to as critical points; crossing one of these points results in a {\em
phase transition} (PT). States lying in different regions generally  have some
strong structural difference and are, in principle,  easily distinguishable  
once somehow a  preferred observable is chosen.

The standard machinery, i.e., the so-called Landau-Ginzburg paradigm, to deal
with this phenomenon is based on the notions  of symmetry breaking, order
parameter, and correlation length \cite{huang}. On the other hand, some system
fails to fall in this conceptual framework. This can be due to the difficulty
of identifying the proper order parameter for systems whose symmetry breaking
pattern is unknown or to the very absence of a local order parameter, e.g.,
quantum phase transitions (QPTs) involving different kinds  of  topological
order \cite{toporder}. Even another standard characterization of QPTs, i.e.,
singularities in the ground state (GS) energy as a function of the  coupling constant, misses
to capture the boundaries between phases for some QPTs, e.g., those with
matrix-product states \cite{wo-ci}.

In the last few years ideas and tools borrowed from quantum information science
\cite{qis} have been used to study quantum, i.e., zero temperature, phase
transistions \cite{sachdev}; in particular the role of quantum
entanglement in QPTs has been extensively investigated \cite{qpt-qis}. More
recently an approach to QPTs based on the concept of {\em quantum fidelity} has
been put forward \cite{za-pa} and applied to systems of quasi-free fermions
\cite{za-co-gio,co-gio-za}, to the so-called matrix-product states
\cite{co-ion-za}, and extended to finite-temperature \cite{zhong-guo}. In the
fidelity approach,  QPTs are identified by studying the behavior of the
amplitude of the overlap, i.e., scalar product, between two ground states
corresponding to two slightly different set of parameters; at QPTs a
drop of the fidelity with scaling behaviour is observed and
quantitative information about critical exponents can be extracted
\cite{co-gio-za,co-ion-za}. The fidelity approach is not based on the
identification of  an order parameter -- and therefore does not require a
knowledge of symmetry breaking patterns -- or more in general on the analysis
of any distinguished observable, e.g., Hamiltonian, but it is a
purely metrical one. All the possible observables are in a sense considered at
once.

In this paper we shall unveil the universal differential-geometric structure 
underlying these observations. We shall show how QPTs can be associated to
the singularities of a Riemannian metric tensor inherited by the
parameter space from the natural Riemannian structure of the
projective space of quantum states.  This structure has an  interpretation in
terms of information-geometry \cite{woo,bra-ca} providing  the
differential-geometric approach of this paper with an information-theoretic
content.

{\em Information-geometry and QPTs}.--
Let us consider a smooth family $H(\lambda),\,\lambda\in{\cal M}$(=the
parameter manifold), of  quantum Hamiltonians in the Hilbert-space $\cal
H$ of the system.
If $|\Psi_0(\lambda)\rangle\in{\cal H}$ denotes the (unique for simplicity)
ground-state of $H(\lambda),$ one has  defined the map $\Psi_0\colon{\cal
M}\rightarrow{\cal H}/\lambda\rightarrow|\Psi_0(\lambda)\rangle$
associating to each set of parameters the ground-state of the corresponding
quantum Hamiltonian. This map can be seen also as a map between ${\cal M}$ and the
projective space $P{\cal H}$(=manifold of ``rays" of $\cal H$). This space  is a
metric space being equipped with the so-called Fubini-Study distance
$d_{FS}(\psi,\phi):= \cos^{-1}{\cal F}(\psi,\phi)$, where
\begin{equation}
{\cal F}(\psi,\phi):=|\langle\psi,\phi\rangle|
\label{fidelity}
\end{equation}
and $\|\psi\|=\|\phi\|=1.$
In Ref. \cite{woo} Wootters showed that this metric has a deep operational
meaning: it quantifies the maximum amount of statistical distinguishability
between the pure quantum states $|\psi\rangle$ and $|\phi\rangle.$ More
precisely, $d_{FS}(\psi,\phi)$ is the maximum over all possible projective
measurements of the Fisher-Rao statistical distance between the probability
distributions obtained from $|\psi\rangle$ and $|\phi\rangle$ \cite{fish}.
Moreover, this result extends to mixed states as well by replacing the
pure-state fidelity (\ref{fidelity}) with the Uhlmann fidelity \cite{Uhlmann}
and the projective measurements with generalized ones \cite{bra-ca}.

These results are non-trivial and allow to identifty in a precise manner the
Hilbert space geometry with a geometry  in the information space:  {\em the
bigger the  Hilbert (or projective) space distance between $|\psi\rangle$ and
$|\phi\rangle$ the higher the degree of statistical distinguishability of
these  two states.}  
From this perspective it is clear that a single real number, i.e., the
distance, virtually encodes information about infinitely many   observables,
e.g., order parameters, one may think to measure. This remark basically
contains the main intuition at the basis of the metric approach to QPTs
advocated in this paper:
at the transition points, a small difference between the
control parameters results in a greatly enhanced distinguishability of the
corresponding GSs, which should be quantitatively revelead by the behavior
of their distance. 

For the purposes of this paper it is crucial to note that the projective  
manifold $P{\cal H}$, besides the structure of metric space,  has a well-known
structure of Riemannian manifold, i.e., it is equipped with a metric tensor.
Here, for the sake of self-consistency, we briefly recall how this Riemannian
metric is obtained starting from the Hilbert space structure of $\cal H.$
$P{\cal H}$ can be seen as the base manifold of a (principal) fiber bundle with
total space given by the unit ball $S$ of $\cal H$, i.e., $S:=\{|\psi\rangle\in
{\cal H}\,/\, ||\psi||=1\}$, and projection $\pi\colon S\rightarrow
P{\cal H}\,/\, |\psi\rangle \rightarrow \{e^{i\theta}|\psi\rangle\,/\,\theta\in
[0,2\pi)\}.$ 
The tangent space to each  point $|\psi\rangle$ of $S$ is isomorphic to a
subspace of $\cal H$ and has therefore defined over it the Hermitean bilinear
form $g_{|\psi\rangle}(u,v):=\langle u, v\rangle$ ($u$ and $v$ are tangent
vectors, i.e., elements of $\cal H$). This defines a (complex) metric tensor
field $g$ over $S.$ To project $g$ down to $P{\cal H}$ one has to introduce the
notion of horizontal subspace for each tangent space of $S$ or equivalently that
of parallel transport and the associated one of connection. In this case the
Hilbert space structure of the tangent spaces provides a natural solution to
this task: the horizontal subspace is simply the set of vectors $|u\rangle$
which are orthogonal to the fiber over $|\psi\rangle$, i.e., $\langle u,
\psi\rangle=0.$ It follows that the complex metric over $P{\cal H}$  is given
by $\tilde{g}_{\pi(|\psi\rangle)}(u,v)=\langle u,
(1-|\psi\rangle\langle\psi|)v\rangle$, called the {\em quantum geometric
tensor} \cite{pro}.
The real (imaginary) part of this
quantity defines  a Riemannian
metric tensor (symplectic form) on $P{\cal H}.$ Another, elementary way of
getting the form of the Riemannian metric over $P{\cal H}$ is by means of Eq.
(\ref{fidelity}). For $\cal F$ very close to the unity, one can write
$d_{FS}^2(\psi,\psi+\delta\psi)\simeq 2(1-{\cal F}).$ Since ${\cal
F}(\psi,\psi+\delta\psi)\simeq |1 +\langle\psi,\delta\psi\rangle +(1/2)
\langle\psi,\delta^2\psi\rangle|^2,$  using this expression and the normalization
of $|\psi\rangle$ one finds
\begin{eqnarray}
ds^2:&=&d_{FS}^2(\psi,\psi+\delta\psi)= \langle\delta\psi,\delta\psi\rangle -|\langle\psi,\delta\psi\rangle|^2
\nonumber\\
&=&\langle \delta\psi,(1-|\psi\rangle\langle\psi|) \delta\psi\rangle \ .
\label{FS}
\end{eqnarray}   

What we would like to do now is to see the metric in the parameter manifold
$\cal M$ induced, i.e., ``pulled-back"  by the ground state mapping $\Psi_0$
introduced above. By writing $\delta|\Psi_0(\lambda)\rangle=\sum_\mu
|\partial_\mu \Psi_0\rangle d\lambda^\mu$,
with $\partial_\mu:=\partial/\partial\lambda^\mu$,
$\mu=1,\ldots,\rm{dim}\,{\cal M}$, and
using Eq. (\ref{FS}), one imediately obtains $ds^2=\sum_{\mu\nu} g_{\mu\nu}
d\lambda^\mu d\lambda^\nu,$ where 
\begin{equation}
g_{\mu\nu}=\Re  \langle\partial_\mu \Psi_0|
\partial_\nu \Psi_0\rangle -\langle \partial_\mu
\Psi_0|\Psi_0\rangle\langle\Psi_0|
\partial_\nu \Psi_0\rangle \ .
\label{g_munu}
\end{equation}
Now we provide a simple perturbative argument on why  one should expect a
singular behavior of the metric tensor at QPTs \cite{BPcomment}. By using
the  first order perturbative expansion
$|\Psi_0(\lambda+\delta\lambda)\rangle\sim |\Psi_0(\lambda)\rangle +\sum_{n\neq
0}(E_0-E_n)^{-1} |\Psi_n(\lambda)\rangle\langle\Psi_n(\lambda)|\delta
H|\Psi_0(\lambda)\rangle$, where $\delta H:=
H(\lambda+\delta\lambda)-H(\lambda)$,
one obtains for the entries of the metric tensor (\ref{g_munu}) the following
expression
\begin{equation}
g_{\mu\nu}= \Re \sum_{n\neq 0}\frac{\langle \Psi_0(\lambda)|\partial_\mu H|
\Psi_n(\lambda)\rangle\langle\Psi_n(\lambda)|\partial_\nu H| \Psi_0(\lambda)\rangle}
{[E_n(\lambda)- E_0(\lambda)]^2} \ .
\label{pert}
\end{equation}
An analogous expression, with the real part replaced by the imaginary one,
gives the antisymmetric tensor which describes the  curvature two-form whose
holonomy is  the Berry phase \cite{BP}.  
Continuous QPTs are  known to occur when, for some specific values of the
parameters and in the thermodynamical limit, the energy gap above the GS
closes. This amounts to a vanishing denominator in Eq. (\ref{pert}) that may
break down the analyticity of the metric tensor entries.

To get further insight about the physical origin of these singularities we
notice that the metric tensor  (\ref{g_munu}) can be cast in an interesting
covariance matrix form \cite{pro}.  In the generic case,  by moving  from $H(\lambda)$ to
$H(\lambda+\delta\lambda)$ no level-crossings occur. In this case  the unitary
operator $O(\lambda,
\delta\lambda):=\sum_n|\Psi_n(\lambda+\delta\lambda\rangle\langle\Psi_n(\lambda)|$
``adiabatically" maps the eigenvectors at $\lambda$ onto those at
$\lambda+\delta\lambda.$ Then by introducing the observables $X_\mu:=
i(\partial_\mu O)O^\dagger$ the metric tensor (\ref{g_munu}) takes the form
$g_{\mu\nu}=(1/2) \langle \{ \bar{X}_\mu ,\, \bar{X}_\nu\} \rangle$
where $\bar{X}_\mu:=X_\mu -\langle X_\mu \rangle.$
Moreover, the line element $ds^2$ can be seen as the variance of the observable
$X:=i(dO)O^\dagger$, i.e., $ds^2=\langle \bar{X}^2 \rangle.$ The operator $X$ is
the  generator of  the map  transforming  eigenstates corresponding to
different values of the parameter into each other. The smaller the difference
between  these eigenstates for a given parameter variation, the smaller the variance of $X.$ Intuitively, at the
QPT one expects to have the maximal possible  difference between
$|\Psi_0(\lambda)\rangle$ and  $|\Psi_0(\lambda+\delta\lambda)\rangle$, i.e.,
many ``unperturbed'' eigenstates $|\Psi_n(\lambda)\rangle$ are needed to
build up the ``new'' GS;
accordingly the variance of $X$ can get very large, possibly divergent. In a
sense  $ds^2$ can be seen as a sort of susceptibility of
the ``order parameter" $X.$

{\em Quasi-Free fermionic systems}.--
In order to show explicitly how the singularities, i.e., divergencies of
$g_{\mu\nu}$ arise, we will discuss the case of the $XY$ model in a detailed
fashion; before doing that we would like to make  some general considerations
about the systems of quasi-free fermions on the basis of the results presented
in Ref. \cite{za-co-gio}.
Systems of quasi-free fermions are defined by the following quadratic
Hamiltonian
\begin{equation}
H=\sum_{i,j=1}^L c_i^\dagger A_{ij} c_j +
\frac{1}{2} \sum_{i,j=1}^L(c_i^\dagger B_{ij} c_j^\dagger +\rm{H.c.}) \ ,
\label{ham}
\end{equation}
where: the $c_i$'s ($c_i^\dagger$'s) are the annihilation (creation) operators
of $L$ fermionic modes, $A,B\in M_L(\mathbb{R})$
are $L\times L$ {\em real}
matrices, symmetric and anti-symmetric respectively, i.e., $A^T=A,\, B^T=-B$.
In Ref. \cite{za-co-gio} it has been shown that the set of GSs of Eq.
(\ref{ham}) is parametrized by orthogonal  $L\times L$ real matrices $T$ giving
the unitary part of the polar decomposition of the matrix $Z:=A-B.$ One can
then prove
that  ${\cal F}(Z,Z^\prime):=|\langle
\Psi_Z|\Psi_{Z^\prime}\rangle|=\sqrt{|\det [(T+T^\prime)/2]|}$
\cite{za-co-gio}. With no loss of generality we can assume $\det(T)=1$  which
identifies the GS manifold of the quasi-free systems (\ref{ham}) with 
$SO(L,\mathbb{R}).$ 
Since $f(Z^\prime):={\cal F}(Z,Z^\prime)$  has a maximum equal to one at
$Z^\prime=Z$ one has $\delta^2 f(Z^\prime)|_Z=\delta^2 \ln
f(Z^\prime)|_Z$; from this,  the expansion for
$Z^\prime\rightarrow Z$ of the above formula for ${\cal F}$ (Eq. (8) in Ref. \cite{za-co-gio}) and by defining
$K:=\ln T \in so(L,\mathbb{R}),$ one finds an explicit form for the metric:
$ds^2\simeq 2(1-{\cal F})=(1/8) {\mathrm{Tr}}(dK)^2.$  From this equation, if
$K=K(\lambda)$, with $\lambda\in{\cal M}$, one obtains the following expression for
the metric tensor induced over $\cal M$, i.e., $g_{\mu\nu}=(1/8) {\mathrm{Tr}} (\partial_\mu
K \partial_\nu K ).$ For translationally invariant Hamiltonians
(\ref{ham}) the anti-symmetric matrix $K$ can be always cast in the canonical
form $K=i\oplus_k \theta_k \sigma_k^y$ where $k$ is a momentum
label. Therefore in this important  case one has
$g_{\mu\nu}=(1/4) \sum_k(\partial\theta_k/\partial\lambda^\mu)
(\partial\theta_k/\partial\lambda^\nu).$ 

We see here that the connection established in Refs.
\cite{za-co-gio,co-gio-za}) between QPTs, e.g., due to  the vanishing of a
quasi-particle energy, and a singularity in the second order expansion of
${\cal F}$ can be directly read  as a {\em connection between QPTs in
quasi-free systems and singularities  in the metric tensor} $g_{\mu\nu}.$ 

The nature of this connection will be now exemplified by considering the QPTs
of the periodic antiferromagnetic $XY$ spin chain in a transverse magnetic
field.
By writing the spin operator in terms of Pauli matrices, i.e.,
$\bm{S}=\bm\sigma/2$, the Hamiltonian for an odd number of spins $L=2M+1$ reads
$H=\sum_{j=-M}^M[-(1+\gamma)\sigma_j^x\sigma_{j+1}^x/4-
(1-\gamma)\sigma_j^y\sigma_{j+1}^y/4+h\sigma_j^z/2]$,
where $\gamma$ is the anisotropy parameter in the $x$-$y$ plane and $h$
is the magnetic field.
This Hamiltonian can be cast in the form (\ref{ham}) by
the Jordan-Wigner transformation.
The critical points of this model are given by the lines
${h}=\pm1$ and by the segment $|{h}|<1,\gamma=0$.
The single particle energies are
$\Lambda_k=\sqrt{\epsilon_k^2+\gamma^2\sin^2(2\pi{}k/L)}$,
where $\epsilon_k=\cos(2\pi{}k/L)-{h}$ and $k=-M,\dots,M$.
For this model the $\theta_k$'s defined above have the form
$\theta_k=\cos^{-1}(\epsilon_k/\Lambda_k)$
and $g_{\mu\nu}=(1/4)\sum_{k=1}^M
(\partial\theta_k/\partial{\lambda^\mu})
(\partial\theta_k/\partial{\lambda^\nu})$,
where $\lambda^{1,2}=h,\gamma$.
One finds $(\partial\theta_k/\partial{h})^2=\gamma^2\sin^2{x_k}/\Lambda_k^4$,
$(\partial\theta_k/\partial\gamma)^2=
\sin^2{x_k}(\cos{x_k}-{h})^2/\Lambda_\nu^4$, and
$(\partial\theta_k/\partial{h})(\partial\theta_k/\partial\gamma)=
\gamma\sin^2{x_k}(\cos{x_k}-{h})/\Lambda_k^4$, with $x_k=2\pi k/L$.

In the thermodynamic limit (TDL), the explicit calculation of $g_{\mu\nu}$ can
be performed analytically.
Indeed, except at critical points, for large $L$ one can replace the discrete
variable $x_k$ with a continuous variable $x$ and substitute the sum with an
integral, i.e., $\sum_{k=1}^M\to[L/(2\pi)]\int_0^\pi\mathrm{d}x$.
At critical points this is not generally feasible due to singularities in some
of the terms in the sums. Outside critical points, the resulting integrals,
albeit non-trivial, yield simple analytical formulas, which differ
depending on whether $|{h}|<1$ or $|{h}|>1$.

\begin{figure}
\includegraphics[width=8.5cm,clip=true,bb=50 50 500 350]{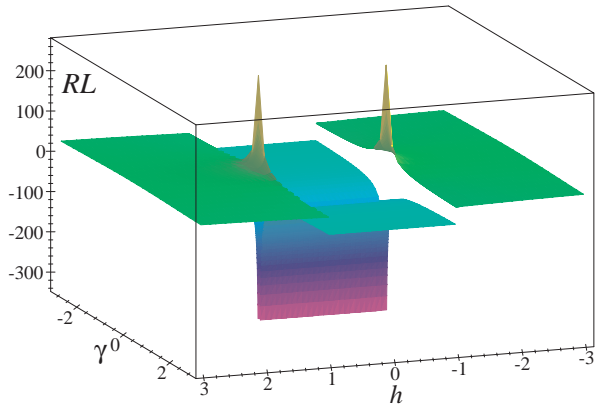}
\caption{Induced curvature $R$ scaled by the system size $L$ for the parameter
space of the $XY$ model.}
\label{fig:1}
\end{figure}

For $|{h}|<1$ in the TDL one finds  a diagonal metric tensor
\begin{equation}
g= \frac{L}{16|\gamma|}\mathrm{diag}
\left(\frac{1}{1-{h}^2},\frac{1}{(1+|\gamma|)^2}\right)
\end{equation}
Closed analytic formulas in the TDL can be obtained also for $|{h}|>1$,
although in a less compact form, which we omit here for brevity.
We only note that for $|{h}|>1$ also the off-diagonal elements of the metric
tensor are non-zero.
Having the induced metric tensor it is also possible to investigate the induced
curvature of the parameter manifold.
We therefore compute the scalar curvature $R$, which is the trace of the Ricci
curvature tensor \cite{Nak}.
We find
$R(|{h}|<1)=-(16/L)(1+|\gamma|)/|\gamma|$ and 
$R(|{h}|>1)=(16/L)(|{h}|+\sqrt{{h}^2+\gamma^2-1})/\sqrt{{h}^2+\gamma^2-1}$.
Note that the curvature diverges on the segment $|{h}|\leq1,\gamma=0$ and is
discontinuous on the lines ${h}=\pm1$. Indeed,
$\lim_{|{h}|\to1^+}R=-\lim_{|{h}|\to1^-}R$.
The behaviour of the curvature $R$ is shown in Fig.~\ref{fig:1}.

{\em Mixed states and classical transitions.}--
In this section we would like to make some extentions of the idea developed in this paper
to finite temperature. This will allow us to establish a connection between the present approach and the one 
for classical PTs developed in \cite{ruppy, brody}.
This latter formalism is in fact obtained in the  special case of commuting density matrices
which effectively turns  the quantum problem into a classical one. 
 
The fidelity approach to QPTs can  be extended to finite-temperature, i.e., to mixed-states, by
using the Uhlmann fidelity \cite{Uhlmann}: 
${\cal F}(\rho_0,\rho_1):= \mathrm{Tr}[\rho_1^{1/2}\rho_0 \rho_1^{1/2}]^{1/2}.$
When $\rho_0$ and $\rho_1$ are commuting operators the fidelity takes the form
${\cal F}(\rho_0,\rho_1)=\sum_n \sqrt{p_n^0 p_n^1}$ where the $p_n^\alpha$ are the eigenvalues of the $\rho_\alpha$'s 
\cite{rel-ent}.
In particular, when 
$\rho_\alpha=Z_\alpha^{-1} \exp(-\beta_\alpha H), Z_\alpha:=\mathrm{Tr}  \exp(-\beta_\alpha H),\,(\alpha=0,1)$
one immediately finds  that the fidelity has a simple expression in terms of partition functioms:
${\cal F}=Z(\beta_0/2+\beta_1/2)(Z(\beta_0) Z(\beta_1))^{-1/2}$ \cite{zhong-guo}.
By expanding for $\beta_0=\beta, \beta_1=\beta+\delta\beta$ one obtains
\begin{equation}
{\cal F}(\beta,\beta+\delta\beta) \simeq
\exp\left[-\frac{\delta\beta^2}{8\beta^2} c_V(\beta) \right] 
\label{fid-thermo}
\end{equation}
where $c_V(\beta)$ denotes the specific heat \cite{huang}.
This relation is remarkable in that it connects the distinguishability degree
of two neighboring thermal quantum states directly to the  macroscopic thermodynamical quantity $c_V.$ 
The line element of the parameter space, i.e., the $\beta$ axis, is then given by 
$ds^2\sim c_V(\beta)\beta^{-2} d\beta^2 
= (\langle H^2\rangle_\beta -\langle H\rangle^2_\beta) d\beta^2.$
A closely related formula has been obtained in \cite{ruppy,brody}. 
Since $PTs$ are associated to anomalies, e.g., divergences, in the behavior of
$c_V(\beta)$, we see that also in this ``classical" case the metric $ds^2$
induced on the parameter space contains signatures of the critical points. In
this sense the information-geometrical approach to PTs seems able to put
quantum and classical PTs under the same conceptual umbrella.

{\em Conclusions.}--
In this paper we proposed a differential-geometric approach to study quantum
phase transtions. The basic idea is that, since distance between quantum states
quantitatively encodes their degree of distinguishability, crossing a critical
point separating regions with structurally different phases  should result in
some sort of singular behaviour  of the metric. This intuition, based on early
studies of quantum fidelity, can be made rigorous in some simple yet important
cases,  e.g., quasi-free fermion systems. The manifold of coupling constants
parameterizing the system's Hamiltonian can be equipped with a (pseudo)
Riemannian tensor $g$ whose singularities correspond to the critical regions.
For the case of the $XY$ chain we explicitely computed the components of $g$ in
the thermodynamic limit, showing that they are
divergent, with universal exponents, at the critical lines. We also computed
the scalar curvature of $g$ and analyzed its relation with criticality. The
geometrical approach advocated in this paper does not depend on the knowledge
of any order parameter or on the analysis of a distinguished observable, it is
universal and information-theoretic in nature.  The study of the physical
meaning of the geometric invariants one can build starting from $g$ (e.g., the
curvature), their finite-size as well as scaling behaviour, and their relations
with the nature  of the quantum phase transition are important questions to be
addressed in future research.

\acknowledgments

%%%%%%%%%%%%%%%%%%%%%%%%%%%%%%%%%%%%%%%%%%%%%%%%%%%%%%%%%%%%%%%%%%%%%%%%%%%%%%%%%%%%%%%%%%%%%%%%%%


\begin{thebibliography}{99}

\bibitem{huang} K.~Huang, \emph{Statistical mechanics}, John Wiley \& Sons, New
York, 1987.

\bibitem{toporder} X.G.~Wen and Q.~Niu, \prb {\bf 41}, 9377 (1990);
X.G.~Wen, \prl {\bf 90}, 016803 (2003).

\bibitem{wo-ci} M.M.~Wolf, G.~Ortiz, F.~Verstraete, and J.I.~Cirac,
Phys. Rev. Lett. {\bf 97}, 110403 (2006).

\bibitem{qis} For a review see, e.g., D.P.~DiVincenzo and C.H.~Bennett,
Nature {\bf 404}, 247 (2000).

\bibitem{sachdev} S.~Sachdev, \textit{Quantum Phase Transitions} (Cambridge
University Press, Cambridge, England, 1999).

\bibitem{qpt-qis} T.J.~Osborne and M.A.~Nielsen, \pra {\bf 66}, 032110 (2002);
A.~Osterloh, L.~Amico, G.~Falci, and R.~Fazio, Nature {\bf 416}, 608 (2002);
G.~Vidal, J.I.~Latorre, E.~Rico, and A.~Kitaev, \prl {\bf 90}, 227902 (2003);
Y.~Chen, P.~Zanardi, Z.D.~Wang, F.C.~Zhang, New J. Phys. {\bf 8}, 97 (2006);
L.-A.~Wu, M.S.~Sarandy, D.A.~Lidar, Phys. Rev. Lett. {\bf 93}, 250404 (2004).

%%%%%%%%%%%%%%%%%%%%%%%%%%%%%%%%%%%%%%%%%%%%%%%%%%%%%%%%%%%%%%%%%%%%%%%%%%%

\bibitem{za-pa} P.~Zanardi and N.~Paunkovic, Phys. Rev. E {\bf 74}, 031123
(2006).
\bibitem{za-co-gio} P.~Zanardi, M.~Cozzini, and P.~Giorda, quant-ph/0606130.
\bibitem{co-gio-za} M.~Cozzini, P.~Giorda, and P.~Zanardi, quant-ph/0608059 (to be
published in Phys. Rev. B).
\bibitem{co-ion-za} M.~Cozzini, R.~Ionicioiu, and P.~Zanardi, cond-mat/0611727.
\bibitem{zhong-guo} P.~Zanardi, H.-T.~Quan, X.-G.~Wang, and C.-P.~Sun,
quant-ph/0612008.

%%%%%%%%%%%%%%%%%%%%%%%%%%%%%%%%%%%%%%%%%%%%%%%%%%%%%%%%%%%%%%%%%%%%%%%%%%%%%%%%%

\bibitem{woo} W.K.~Wootters, Phys. Rev. D, {\bf 23}, 357 (1981).

\bibitem{bra-ca}  S.L.~Braunstein and C.M.~Caves, Phys. Rev. Lett. {\bf 72}, 3439
(1994).

\bibitem{fish} Given the complete set of one-dimensional projections
$\{|i\rangle\langle i|\}_{i=1}^{\rm{dim}\,{\cal H}}$ one has the two probability
distributions $p_i=|\langle i|\psi\rangle|^2$ and $q_i=|\langle
i|\phi\rangle|^2.$ For $p$ and $q$ infintesimally close to each other the Fisher
distance is given by $d_F(p,q):=\sum_i (p_i-q_i)^2/p_i= \sum_i p_i (d\ln p_i)^2.$

\bibitem{Uhlmann} A.~Uhlmann, Rep. Math. Phys. {\bf 9}, 273 (1976);
R.~Jozsa, J. Mod. Opt. {\bf 41}, 2315 (1994).

\bibitem{pro} J.P.~Provost and G.~Vallee, Commun. Math. Phys. {\bf 76}, 289
(1980).

\bibitem{BP} For a reprint collection see {\em Geometric phases in Physics},
A.~Shapere and F.~Wilczek (Eds), World Scientific, Singapore, 1889.

\bibitem{BPcomment} 
This heuristic argument is going to be parallel to the one used in
Ref.~\cite{BP-qpt} where the relation between QPTs an Berry phases is studied.
While the Berry curvature can be identically vanishing, e.g., for real
wavefunctions, this cannot be the case for the quantity (\ref{g_munu}).

\bibitem{BP-qpt}
S.-L.~Zhu, Phys. Rev. Lett. {\bf 96}, 077206 (2006);
A.~Hamma, quant-ph/0602091.

\bibitem{ruppy} G.~Ruppeiner, Phys. Rev. A {\bf 20}, 1608 (1979);
Rev. Mod. Phys. {\bf 67}, 605 (1995) and references therein.

\bibitem{brody}
D.~Brody and N.~Rivier,
Phys. Rev. E {\bf 51}, 1006 (1995).

\bibitem{Nak} See, for example, M. Nakahara,
{\em Geometry, topology and Phsyics}, Institute of Physics Publishing (1990).

\bibitem{rel-ent}
For $p_n^1=p_n^0+dp_n$ ($dp_n\approx0$) we have that ${\cal
F}(\rho_0,\rho_1)\approx 1-(1/8)\sum_n dp_n^2/p_n^0$; thus, to lowest non-zero
order, the difference of the fidelity from 1 is proportional to the
Fisher-Rao distance, that, for infinitesimal variations, coincides with the
relative entropy between the two probability distributions $\{p_n^0\}$ and
$\{p_n^1\}$.

\end{thebibliography}
\end{document}